# Magnetic and magnetotransport properties of ThCr$_2$Si$_2$-type Ce$_2$O$_2$Bi composed of conducting Bi$^{2-}$ square net and magnetic Ce−O layer

Shunsuke Shibata[1], Ryosuke Sei[1,2], Tomoteru Fukumura[2,3,a)], Tetsuya Hasegawa[1]

[1]*Department of Chemistry, The University of Tokyo, Tokyo 113-0033, Japan*

[2]*Department of Chemistry, Tohoku University, Sendai 980-8578, Japan*

[3]*WPI Advanced Institute for Materials Research, Tohoku University, Sendai 980-8577, Japan*

ThCr$_2$Si$_2$-type Ce$_2$O$_2$Bi epitaxial thin film was grown by multilayer solid phase epitaxy recently developed. Ionic state of Ce was confirmed to be 3+ by x-ray photoelectron spectroscopy, corresponding to electronic configuration of [Xe]4f$^1$. Electrical resistivity showed nonmonotonic temperature dependence with a sharp resistivity maximum, concomitant with a magnetization kink at 10 K suggesting antiferromagnetic transition. In addition, magnetoresistance showed a large angular-dependent magnetoresistance. These results imply that carrier transport in Bi$^{2-}$ square net could be influenced by magnetic ordering in Ce−O layer owing to its unique layered structure [Bi$^{2-}$ / (Ce$_2$O$_2$)$^{2+}$], particularly in the form of epitaxial thin film.

[a)] Electronic mail: tf1@tohoku.ac.jp

ThCr$_2$Si$_2$-type $R_2$O$_2$Bi ($R$ = Y or rare earth elements) compounds have a layered structure composed of Bi$^{2-}$ square net and [$R_2^{3+}$O$_2^{2-}$]$^{2+}$ layers exhibiting various electric and magnetic properties depending on $R$ cation,[1] in addition to recently discovered superconductivity in Y$_2$O$_2$Bi.[2] A metal-to-insulator transition was reported from $R$ = Y to La owing to the chemical pressure.[1] The band calculation studies showed that the Fermi surface is composed of Bi 6p$_{x,y}$ bands in $R_2$O$_2$Bi ($R$ = La,[1,3] Er[3]) irrespective of the element $R$. Y$_2$O$_2$Bi and La$_2$O$_2$Bi without f electron showed temperature independent paramagnetic susceptibility, while Pr$_2$O$_2$Bi, Gd$_2$O$_2$Bi, and Er$_2$O$_2$Bi with 4f$^n$ electron configuration were reported to show antiferromagnetic transition.[1] All those samples were polycrystalline powders, thus their single crystalline samples are desired to further investigate the intrinsic properties. Recently, we developed two types of solid phase epitaxy for thin film growth of Y$_2$O$_2$Bi, using powder precursor with two step heating[4] and multilayer precursor with in situ one step heating.[6] In the latter method, high quality Y$_2$O$_2$Bi epitaxial thin film was obtained, resulting in observation of two dimensional weak antilocalization effect in Bi$^{2-}$ square net layer. This method can be applied to epitaxial thin film growth of various $R_2$O$_2$Bi. In addition, the epitaxial stabilization of chemically unstable phase is expected.

Polycrystalline Ce$_2$O$_2$Bi has been synthesized previously, but its physical properties have been scarcely reported probably because of its chemical instability in air.[1,7,8] The valence of Ce cations is supposed to be 3+ with electronic configuration of [Xe]4f$^1$, hence the antiferromagnetic ordering is expected as described above. Accordingly, the interplay between antiferromagnetic Ce−O layer and electrically conducting two dimensional Bi$^{2-}$ square net can be investigated, particularly in case of epitaxial thin film. In this paper, we report on the growth of Ce$_2$O$_2$Bi (001) epitaxial thin film by multilayer solid phase epitaxy and the electrical and magnetic properties. A sharp resistivity maximum concomitant with a magnetization kink at 10 K suggesting antiferromagnetic transition and a large angular-dependent magnetoresistance were observed, implying a significant effect of magnetic ordering in Ce−O layer on two dimensional electronic transport in Bi$^{2-}$ square net.

Multilayer solid phase epitaxy was applied to grow Ce$_2$O$_2$Bi epitaxial thin film. Firstly, a multilayer precursor [Bi (1.6 nm) / Ce (1.7 nm) / Bi (1.6 nm) / Ce (1.7 nm) / CeO$_2$ (1.1 nm)]$_{25}$ capped with Ce (2.6 nm) layer was deposited on CaF$_2$ (100) single-crystal substrate ($a/\sqrt{2}$ = 0.387 nm) or SrF$_2$ (100) single-crystal substrate ($a/\sqrt{2}$ = 0.410 nm) in Ar gas pressure of $2 \times 10^{-2}$ Torr at room temperature by DC and RF sputtering (Figure 1(a)). Sputtering targets were Ce (purity of 99.9%), CeO$_2$ (99.9%), and Bi (99.995%). The multilayer precursor was in situ heated in Ar gas at 850 °C for 10 minutes to form Ce$_2$O$_2$Bi epitaxial thin film via solid phase epitaxy. The crystal structure was determined by x-ray diffraction method (XRD) with Cu Kα radiation equipped with one and two dimensional detectors (D8 DISCOVER, Bruker AXS). The film thickness was typically 100 nm. X-ray photoelectron spectroscopy (XPS) of Ce 3d was measured in ultrahigh vacuum after surface cleaning by in situ Ar sputtering (PHI 5000 VersaProbe, ULVAC-PHI). The XPS spectra were calibrated by using the value of C 1s peak (248.8 eV). Electrical transport properties were measured with standard four-probe method by physical property measurement system equipped with sample rotator system (PPMS, Quantum Design). Magnetization was measured by superconducting quantum interference device magnetometer (MPMS, Quantum Design). The film on $SrF_2$ was used for measurements of XRD, XPS, and electric transport properties, and the film on $CaF_2$ was used for measurements of magnetization.[5] $Y_2O_2Bi$ (001) epitaxial thin film on $CaF_2$ substrate was prepared and the electric and magnetic properties were measured for reference.[6]

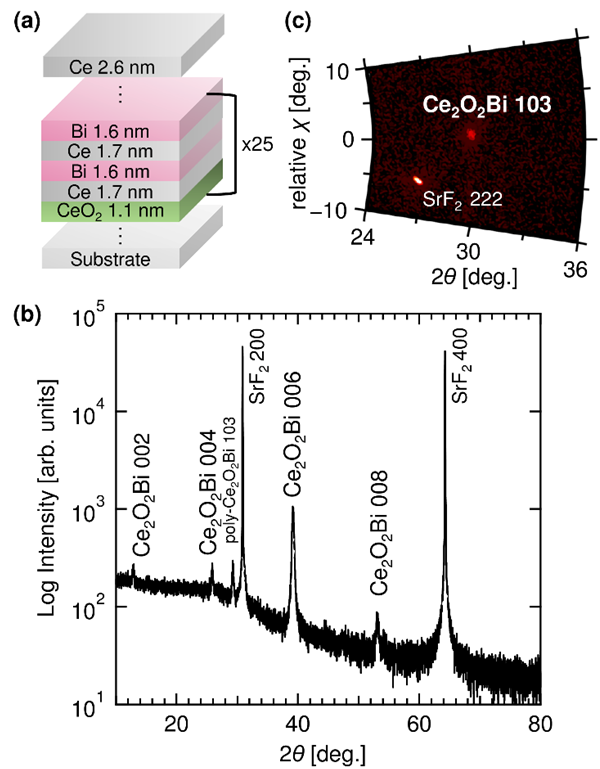

FIG. 1. (a) Schematic multilayer precursor to form $Ce_2O_2Bi$ epitaxial thin film by multilayer solid phase epitaxy. (b) $\theta$–$2\theta$ XRD pattern for $Ce_2O_2Bi$ (001) epitaxial film on $SrF_2$ (001) substrate. (c) Two-dimensional XRD pattern of asymmetric lattice plane at $\chi = 41°$ for $Ce_2O_2Bi$ (001) epitaxial thin film on $SrF_2$ (001) substrate.

Figure 1(b) shows the out-of-plane XRD pattern of $Ce_2O_2Bi$ thin film on $SrF_2$ (100) substrate. $Ce_2O_2Bi$ 00$n$ ($n$ = 2, 4, 6, 8) peaks were clearly observed, in addition to small $Ce_2O_2Bi$ 103 peak corresponding to the polycrystalline phase confirmed by two dimensional XRD pattern. The volume fraction of the polycrystal was estimated to be approximately 30% from the XRD areal peak intensity. Figure 1(c) shows intense $Ce_2O_2Bi$ 103 spot in two dimensional XRD pattern of asymmetric plane at $\chi = 41°$. From these results, the film was mainly composed of $Ce_2O_2Bi$ (001) epitaxial thin film mixed with the small amount of the polycrystalline $Ce_2O_2Bi$.[9] Similar result was obtained for $Ce_2O_2Bi$ thin film on $CaF_2$ (100) substrate. The *a*- and *c*- axis lengths of films were 0.408 nm and 1.38 nm on $SrF_2$ substrate and 0.399 nm and 1.40 nm on $CaF_2$ substrate, respectively. In case of bulk polycrystal, the lattice constants were reported to be $a$ = 0.40369(5) nm and $c$ = 1.3746(2) nm.[8] In comparison with the bulk lattice constants, *a*- and *b*- axes of the film on $SrF_2$ increased under tensile strain with in-plane lattice mismatch of −1.8%, while those of the film on $CaF_2$ decreased under compressive strain with in-plane lattice mismatch of 4.2%. This rather larger mismatch than that of 0.23% in case of $Y_2O_2Bi$ film on $CaF_2$ (Ref.4) might degrade the crystallinity as exemplified by the presence of polycrystalline phase. $Ce_2O_2Bi$ films were decomposed in air, but the slow decomposition speed enabled to measure various physical properties as described below.

Figure 2 shows the Ce 3*d* XPS spectrum for $Ce_2O_2Bi$ film. The spectra for $Ce_2O_3$ and $CeO_2$ epitaxial thin films are also shown for reference.[10] XPS spectrum of $Ce_2O_3$ shows four peaks corresponding to the electronic state of $Ce^{3+}$,[10−13] whose peak position was completely different from those of $CeO_2$. The spectral shape and the peak position show significant resemblance between $Ce_2O_2Bi$ and $Ce_2O_3$, representing the $Ce^{3+}$ state in $Ce_2O_2Bi$ similar to other $R_2O_2Bi$ compounds.[1]

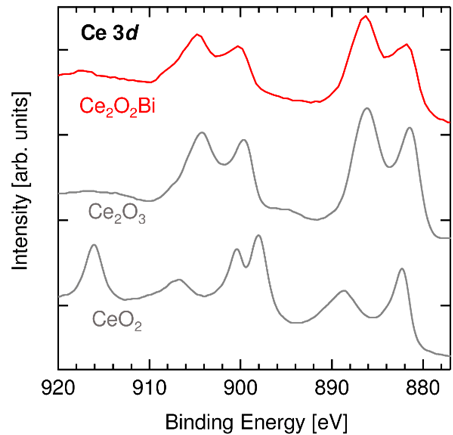

FIG. 2. Ce 3*d* XPS spectrum for $Ce_2O_2Bi$ epitaxial thin film measured after surface sputtering. XPS spectra of $Ce_2O_3$ and $CeO_2$ films are shown for a reference.[10] Adapted with permission from Ref. 10. Copyright (2013) American Chemical Society.

Figure 3(a) shows temperature dependence of magnetization for $Ce_2O_2Bi$ film under field cooling at 0.05 T. The magnetization increased with decreasing temperature, showing a small kink at 10 K. Such kink structure was also observed in $Pr_2O_2Bi$, $Gd_2O_2Bi$, and $Er_2O_2Bi$ polycrystalline powders at 15.0 K, 10.1 K, and 3.0 K, respectively, attributed to antiferromagnetic transition.[1] The small magnetic hysteresis at 2 K also suggests the antiferromagnetic order in $Ce_2O_2Bi$ (inset of Figure 3(b)). The saturation magnetization of $Ce_2O_2Bi$ film, about 1 $\mu_B$ / Ce cation at 2 K, was much larger than that of $Y_2O_2Bi$ film (Figure 3(b)).[6] This result reflects a principal role of $4f^1$ electron in magnetism of Ce−O layer, being consistent with the $Ce^{3+}$ state observed by XPS measurement. This almost fully saturated $Ce^{3+}$ moments suggest a metamagnetic transition possibly due to weak antiferromagnetic exchange coupling in quasi-two dimensional Ce−O layer, although more precise magnetic structure has to be clarified by using neutron scattering measurement, for example.

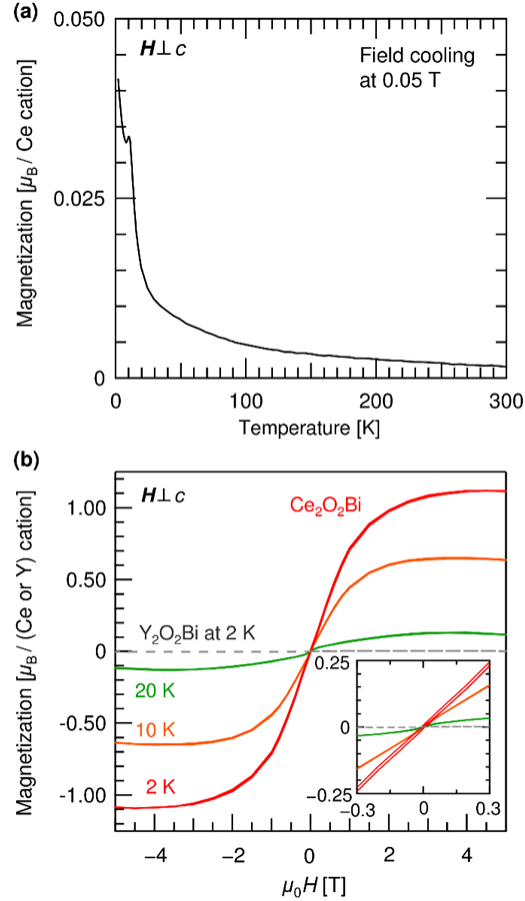

FIG. 3. (a) $M-T$ curve of $Ce_2O_2Bi$ epitaxial thin film with field cooling at 0.05 T. The magnetic field was applied along $ab$-plane. (b) $M-H$ curves of $Ce_2O_2Bi$ epitaxial thin film at 2 K, 10 K, and 20 K. The gray curve denotes $M-H$ curve of $Y_2O_2Bi$ epitaxial thin film at 2 K. Inset shows a magnified view.

Figure 4 shows $\rho-T$ curves of $Ce_2O_2Bi$ film under out-of-plane magnetic field from 0 T to 9 T with step of 1 T. In contrast with $Y_2O_2Bi$ film, the resistivity showed a significant increase with decreasing temperature and a sharp resistivity maximum at about 14 K at 0 T, showing good coincidence with the magnetization kink at 10 K (see Figure 3(a)). With further decreasing temperature down to 2 K, the resistivity showed a sharp drop. Such abrupt increase and sharp drop in $\rho$ have not been observed in any $R_2O_2Bi$ polycrystals and $Y_2O_2Bi$ film.[1,6] Suppressed resistivity peak with the application of magnetic field indicates that the resistivity peak was caused by spin scattering at magnetic transition as have been often observed in various magnetic conductors. Here, it is noted that localized spins and conducting carriers in $Ce_2O_2Bi$ are separately confined in Ce−O layer and $Bi^{2-}$ square net, respectively, implying a significant interplay between magnetic Ce−O layer and adjacent two dimensional $Bi^{2-}$ square net.

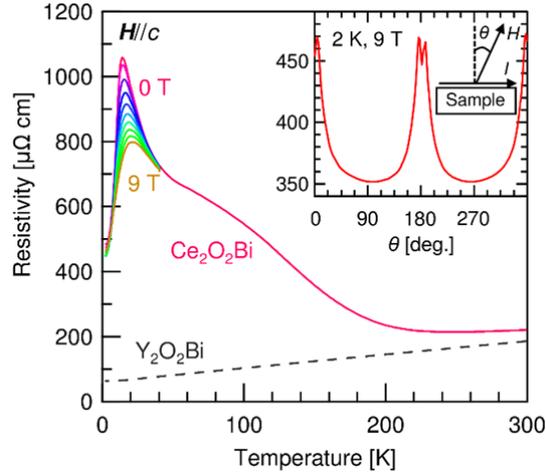

FIG. 4. $\rho$−$T$ curves at 0−9 T along surface normal with step of 1 T for $Ce_2O_2Bi$ and $Y_2O_2Bi$ (Ref.6) (gray curve) epitaxial thin films. The inset shows angular dependence of magnetoresistance at 2 K and 9 T for $Ce_2O_2Bi$ epitaxial thin films and the measurement geometry, in which $\theta$ is angle between surface normal to magnetic field.

The inset of Figure 4 shows angular dependence of magnetoresistance at 2 K and 9 T. The angular dependence showed a sharp maximum at out-of-plane magnetic field and a broad minimum at in-plane magnetic field, where the small double splitting at the maximum might be caused by sample imperfection because of its sample-dependent appearance. The angular dependence of magnetoresistance, that was much larger than that of $Y_2O_2Bi$,[6] was comparable to those of $SrZnMn_2$-type $CaMnBi_2$ and $SrMnBi_2$ single crystals with similar Bi square net structure,[14,15] in spite of the dissimilar angular dependence: the sharp change at around out-of-plane magnetic field for $Ce_2O_2Bi$ while at in-plane magnetic field for $CaMnBi_2$ and $SrMnBi_2$. The angular dependence might be associated with magnetic anisotropy of quasi-two dimensional Ce−O layers interacting with conduction carriers in $Bi^{2-}$ square net. The crystal structure of $Ce_2O_2Bi$ could be regarded as infinite-layered [metal / magnetic insulator] structure. Thus, it will be intriguing to investigate spin filter tunneling phenomena as observed in [metal / ferromagnetic insulator / metal] trilayer junction.[16-18]

In conclusion, we grew $Ce_2O_2Bi$ epitaxial thin film by using multilayer solid phase epitaxy. $Ce_2O_2Bi$ showed a sharp resistivity maximum almost concomitant with the possible antiferromagnetic transition at 10 K. The peculiar magnetoresistance suggests that the layered structure in $Ce_2O_2Bi$, composed of two dimensionally conducting $Bi^{2-}$ square net and antiferromagnetic Ce−O layer, is a good platform to investigate the interplay between these layers such as spin filter tunneling.

XPS measurement was conducted in Research Hub for Advanced Nano Characterization, The University of Tokyo, under the support of Nanotechnology Platform by MEXT, Japan (No. 12024046). This research was in part supported by JSPS KAKENHI (26600091, 26105002) and CREST, JST.

[1] H. Mizoguchi and H. Hosono, J. Am. Chem. Soc. **133**, 2394 (2011).

[2] R. Sei, S. Kitani, T. Fukumura, H. Kawaji, and T. Hasegawa, J. Am. Chem. Soc. **138**, 11085 (2016).

[3] H. Kim, C.-J. Kang, K. Kim, J. H. Shim, B. I. Min, Phys. Rev. **B 93**, 125116 (2016).

[4] R. Sei, T. Fukumura, and T. Hasegawa, Cryst. Growth Des. **14**, 4227 (2014).

[5] Effects of the strain from these substrates on the properties of the flims were not observed possibly due to the insufficient crystallinity and/or the chemical degradation.

[6] R. Sei, T. Fukumura, and T. Hasegawa, ACS Appl. Mater. Interfaces **7**, 24998 (2015).

[7] R. Benz, Acta Crystallogr. Sect. B **27**, 853 (1971).

[8] J. Nuss and M. Jansen, J. Alloys Cmpd. **480**, 57 (2009).

[9] All the XRD peaks disappeared within several days in air due to the chemical degradation.

[10] V. Stetsovych, F. Pagliuca, F. Dvořák, T. Duchoň, M. Vorokhta, M. Aulická, J. Lachnitt, S. Schernich, I. Matolínová, K. Veltruská, T. Skála, D. Mazur, J. Mysliveček, J. Libuda, V. Matolín, J. Phys. Chem. Lett. **4**, 866 (2013).

[11] A. Pfau and K. D. Schierbaum, Surf. Sci. **321**, 71 (1994).

[12] W. Xiao, Q. Guo, and E. G. Wang, Chem. Phys. Lett. **368**, 527 (2003).

[13] T. Nakano, A. Kotani, and J. C. Parlebas, J. Phys. Soc. Jpn. **56**, 2201 (1987).

[14] K. Wang, D. Graf, H. Lei, S. W. Tozer, and C. Petrovic, Phys. Rev. B **84**, 220401 (2011).

[15] K. Wang, D. Graf, L. Wang, H. Lei, S. W. Tozer, and C. Petrovic, Phys. Rev. B **85**, 041101 (2012).

[16] L. Esaki, P. J. Stiles, and S. von Molnar, Phys. Rev. Lett. **19**, 852 (1967).

[17] J. S. Moodera, X. Hao, G. A. Gibson, and R. Meservey, Phys. Rev. Lett. **61**, 637 (1988).

[18] M. Gajek, M. Bibes, A. Barthélémy, K. Bouzehouane, S. Fusil, M. Varela, J. Fontcuberta, and A. Fert, Phys. Rev. B **72**, 020406 (2005).